\input{aipcheck}

\documentclass[
    ,final            
  ]
  {aipproc}

\layoutstyle{6x9}

\usepackage{amssymb,amsmath,array}
\usepackage{wrapfig,rotating}

\setcitestyle{numbers}

\begin{document}

FERMILAB-CONF-11-344-E

\title{$D^{*}$ production in deep-inelastic Scattering\\ at low $Q^2$}


\author{Andreas W. Jung~~(for the H1 collaboration)\\}{
        address={Fermi National Accelerator Laboratory (Fermilab)\\
        E-mail: ajung@fnal.gov}
}



\begin{abstract}
Inclusive production of $D^*$ mesons in deep-inelastic scattering at HERA
is studied in the range $5<Q^2<100~\mathrm{GeV^2}$ of the photon virtuality
and $0.02<y<0.70$ of the inelasticity of the scattering process. The visible
range for the $D^*$ meson is $p_{_{T}}(D^*) > 1.25~\mathrm{GeV}$ and
$|\eta (D^*)| < 1.8$. The data were taken with the H1 detector in the
years 2004 to 2007 and correspond to an integrated luminosity
of $~347~\mathrm{pb^{-1}}$. Single and double differential cross
sections are measured. The results are compared to QCD predictions.
\end{abstract}

\maketitle


\section{Introduction}
HERA was the unique electron-proton ($ep$) accelerator colliding $27.6~\mathrm{GeV}$ electrons (positrons)
 with $920~\mathrm{GeV}$ protons providing a center-of-mass energy of $\sqrt{s} = 318~\mathrm{GeV}$.
The charm quark production in $ep$ scattering is dominated by the boson-gluon-fusion (BGF) process 
$(\gamma p \rightarrow c\bar{c})$. This production process is directly sensitive to the gluon density 
in the proton and allows its universality to be tested. The kinematic region of deep-inelastic scattering 
(DIS) is defined by the four-momentum transfer squared ($Q^2$) of the 
exchanged photon: $Q^2 \gtrsim 5~\mathrm{GeV^{2}}$. Due to the presence of a hard scale ($m_{c}, Q^2$ or $p_{_{T}}$) 
perturbative Quantum Chromodynamics (pQCD) can be applied. If one of the other scales is much bigger than the mass, 
charm quarks can be treated as massless ("massless scheme"), otherwise the mass needs to be taken into account 
("massive scheme"). The latter assumes no charm quark content of the proton. 

\section{$\mathbf{D^{*\pm}}$ Cross Section Measurements in DIS}
\begin{figure}[ht]
    \centerline{\includegraphics[width=0.85\columnwidth]{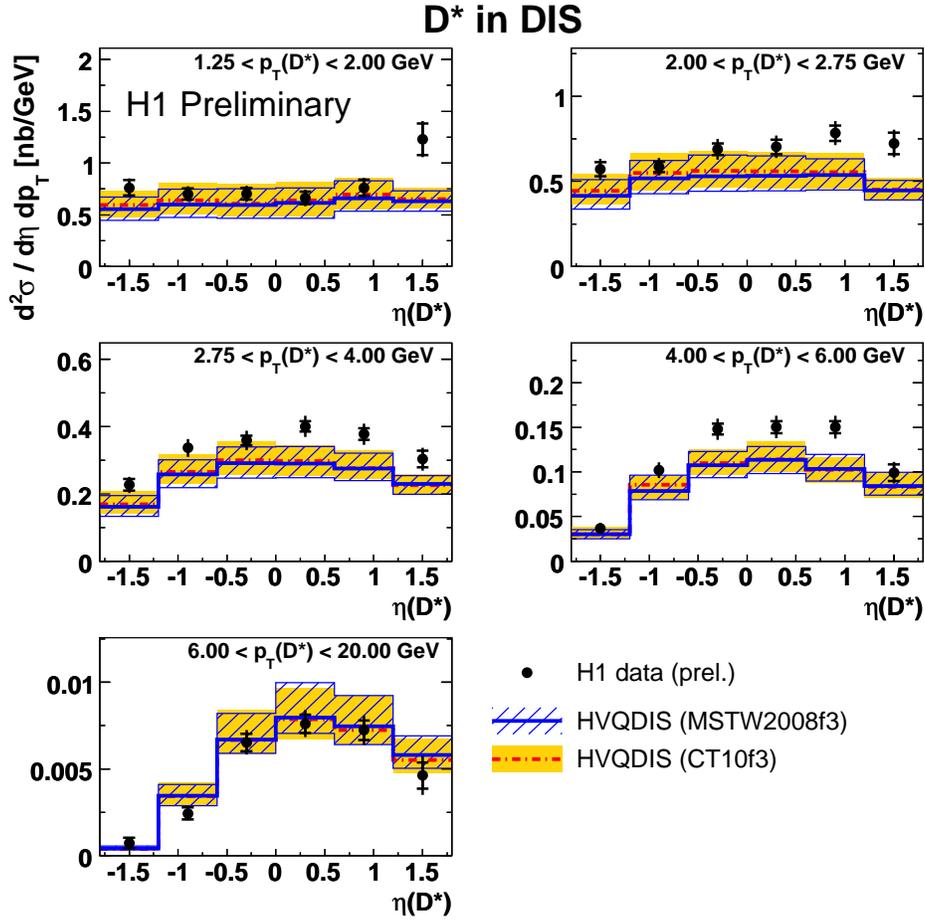}}
  \caption{\label{fig:xsec_2d} Two dimensional $D^*$ cross section as a function of $\eta(D^*)$ and $p_{_{T}}(D^*)$ compared to the massive NLO QCD calculation (HVQDIS) using two different proton PDFs: MRST2008f3 or CT10f3.}
 \end{figure}

\begin{figure}[ht]
    \begin{minipage}[c]{0.475\columnwidth}
   \centerline{\includegraphics[width=0.95\columnwidth]{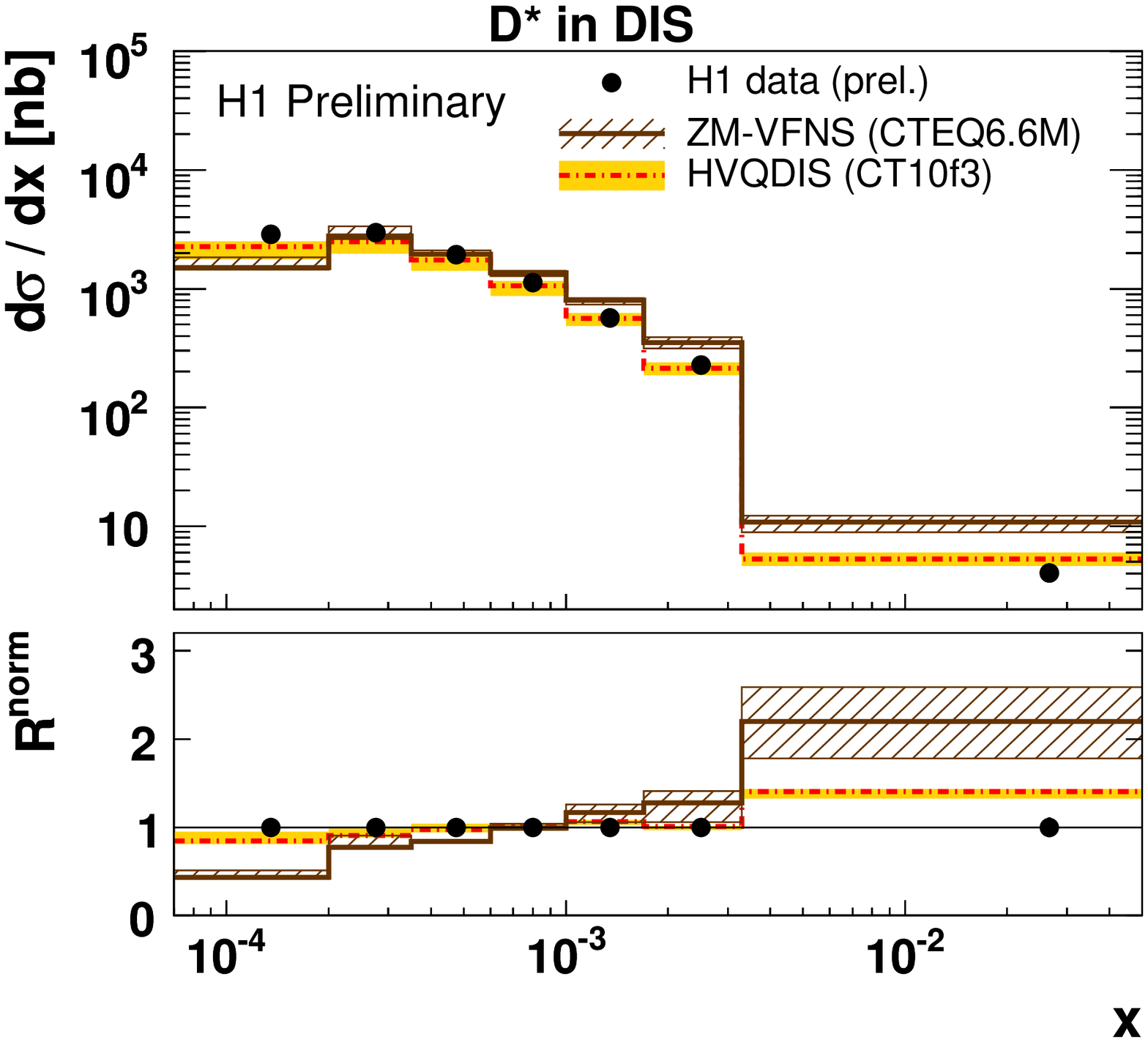}}
     \end{minipage}
     \hspace{.05\linewidth}
     \begin{minipage}[c]{0.475\columnwidth}
    \centerline{\includegraphics[width=0.95\columnwidth]{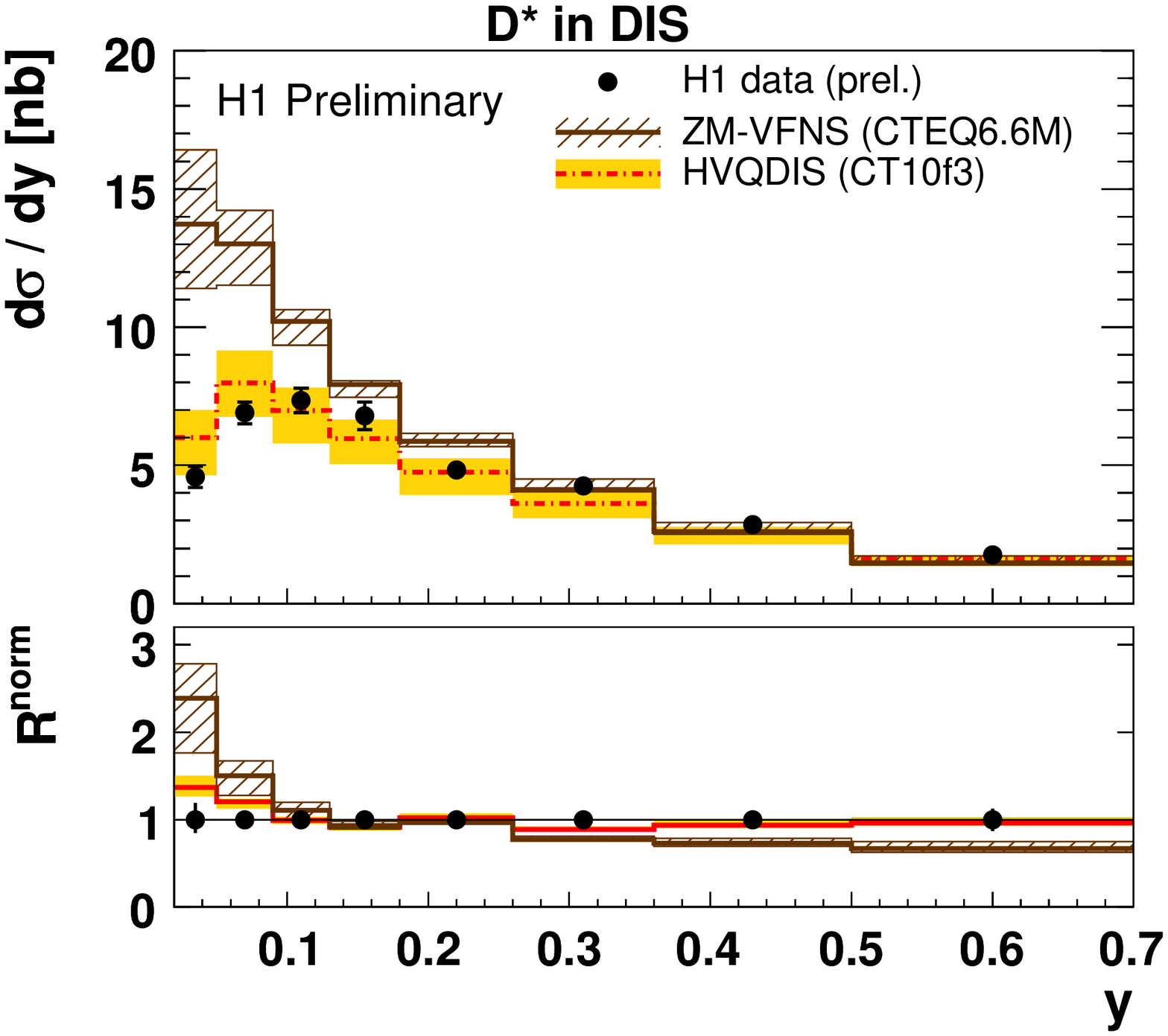}}
     \end{minipage}
  \caption{\label{fig:xsec_x_zm} $D^*$ cross section as a function of $x$ (left) and $y$ (right) for $p_{_{T}}^* > 2~\mathrm{GeV}$ compared 
to the massive NLO QCD calculation (HVQDIS) using the proton PDF: CT10f3 and to the massless (ZM-VFNS) NLO QCD calculation using CTEQ6.6M \cite{cteq66}.}    
 \end{figure}

\begin{figure}[ht]
    \centerline{\includegraphics[width=0.55\columnwidth]{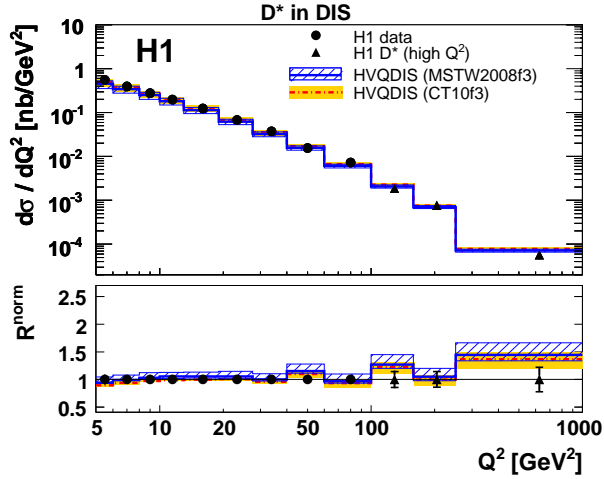}}
  \caption{\label{fig:allq2} $D^*$ cross section as a function of $Q^2$ with the additional cuts: $|\eta(D^*)| < 1.5$ and $p_{_{T}}(D^*)>1.5~\mathrm{GeV}$ are shown together with the results from the high $Q^2$ $D^*$ publication \cite{medQ2Prel}. Data are compared to the massive NLO QCD calculation (HVQDIS) using two different proton PDFs: MRST2008f3 or CT10f3.}
 \end{figure}

Events containing charm quarks are efficiently identified reconstructing $D^*$ mesons by the mass difference method. 
$D^*$ production in DIS has been measured for photon virtualities of: $5<Q^2<100~\mathrm{GeV^2}$ and inelasticities 
of: $0.02<y<0.70$. The full HERAII data set corresponding to an integrated luminosity of $347~\mathrm{pb^{-1}}$ has been
analyzed. The visible phase space of the $D^*$ meson is restricted to $p_{_{T}} (D^*) > \mbox{1.25}~\mathrm{GeV}$ 
and $|\eta (D^*)| < \mbox{1.8}$ \cite{medQ2Prel}. This measurement yields currently the largest phase space coverage 
at HERA for an inclusive $D^*$ cross section measurement.\\
\begin{figure}[ht]
    \begin{minipage}[c]{0.475\columnwidth}
    \centerline{\includegraphics[width=0.95\columnwidth]{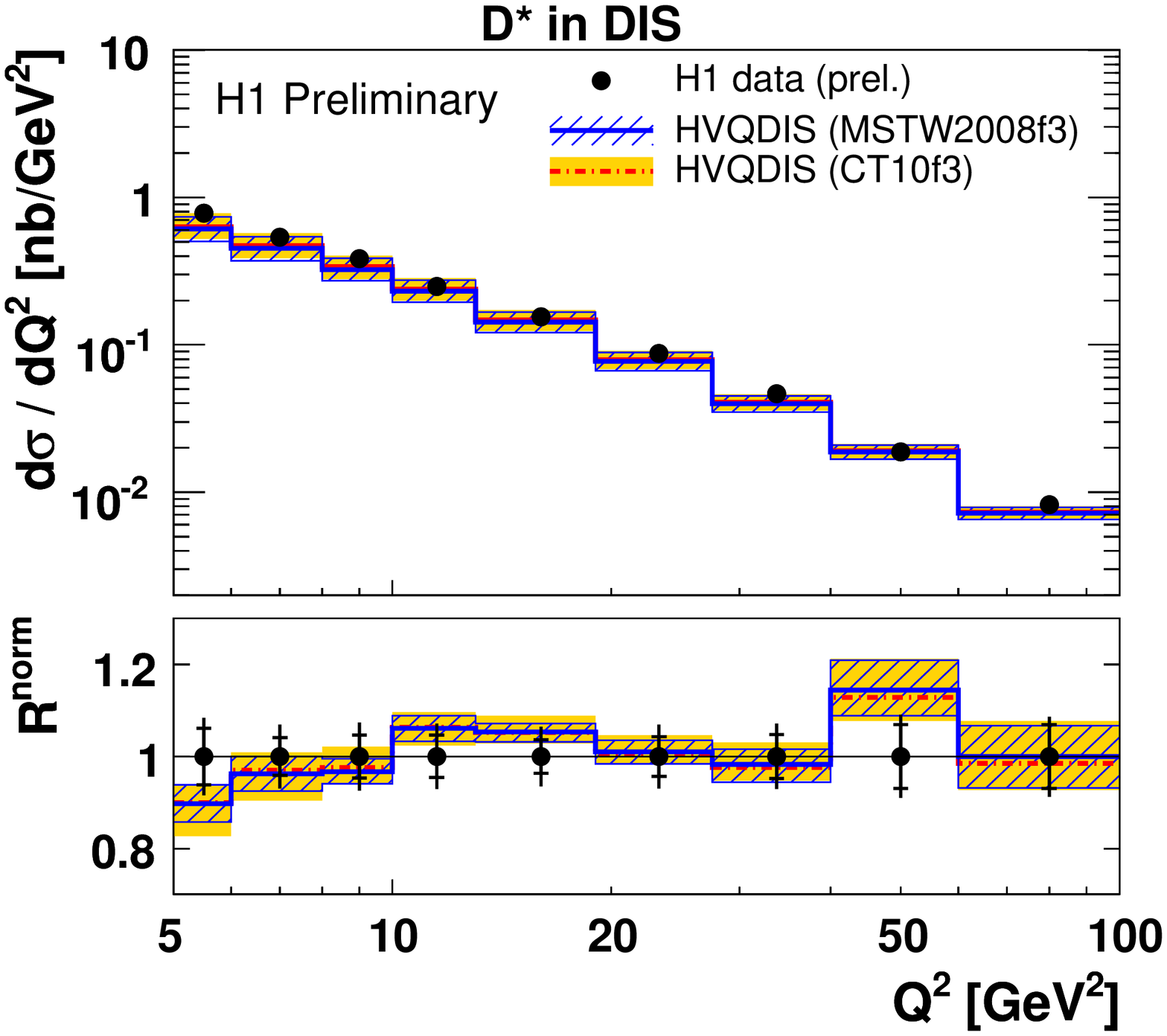}}
     \end{minipage}
     \hspace{.05\linewidth}
     \begin{minipage}[c]{0.475\columnwidth}
    \centerline{\includegraphics[width=0.95\columnwidth]{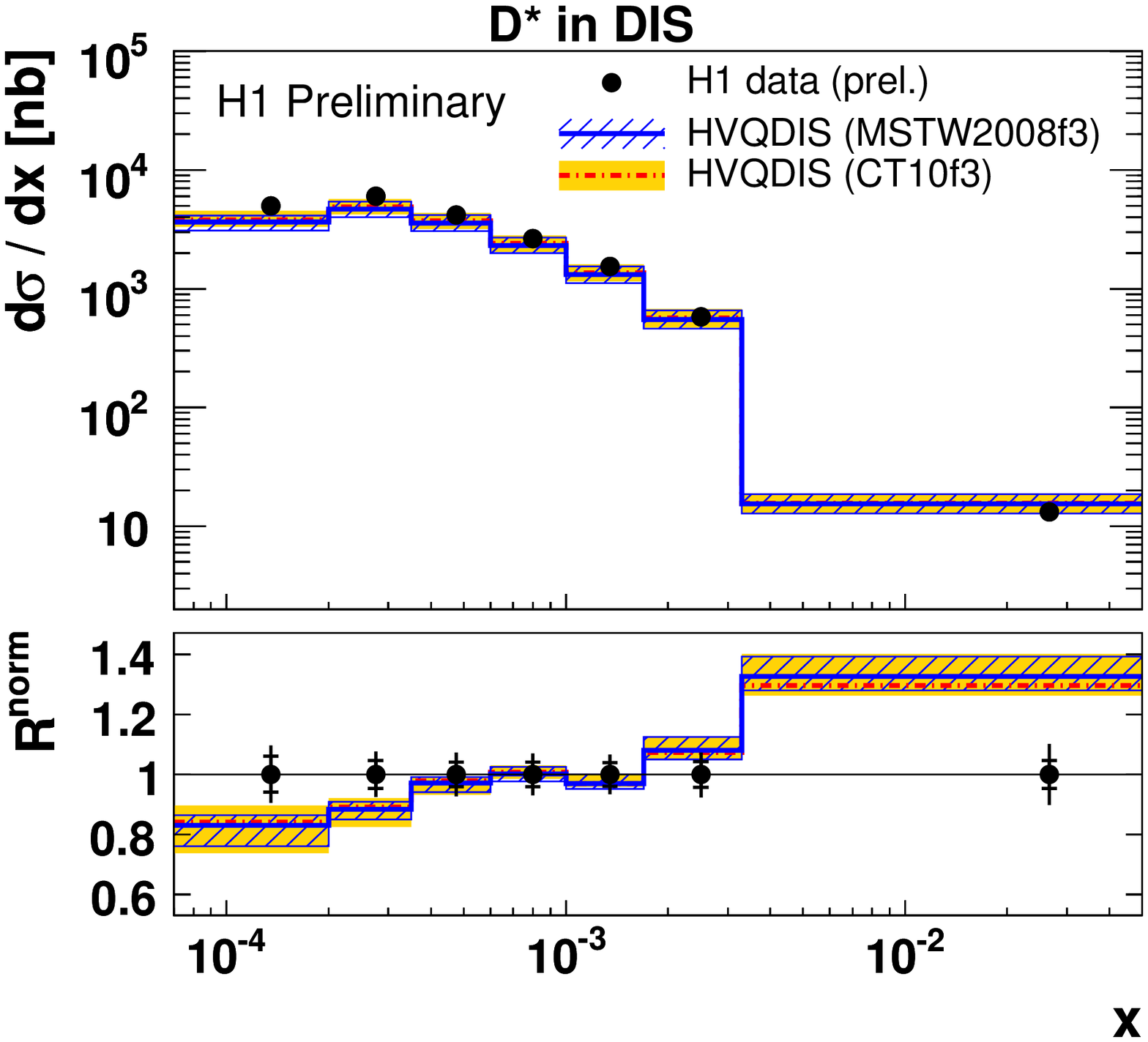}}
     \end{minipage}
  \caption{\label{fig:xsec_q2_x} $D^*$ cross section as a function of $Q^2$ (left) and $x$ (right) compared to the massive NLO QCD 
calculation (HVQDIS) using two different proton PDFs: MRST2008f3 \cite{mrst08f3} or CT10f3 \cite{ct10f3}.}
 \end{figure}
The $D^*$ cross section as a function of $Q^2$ and $x$ is depicted in Figure \ref{fig:xsec_q2_x}(left) and (right). The data 
are compared to the massive NLO QCD calculation (HVQDIS) using two different proton parton density functions (PDFs): MRST2008f3 \cite{mrst08f3} or CT10f3 \cite{ct10f3}. The $Q^2$ dependence of the data is well described, while the slope in $x$ is not very well reproduced.
\begin{figure}[ht]
    \begin{minipage}[c]{0.475\columnwidth}
    \centerline{\includegraphics[width=0.95\columnwidth]{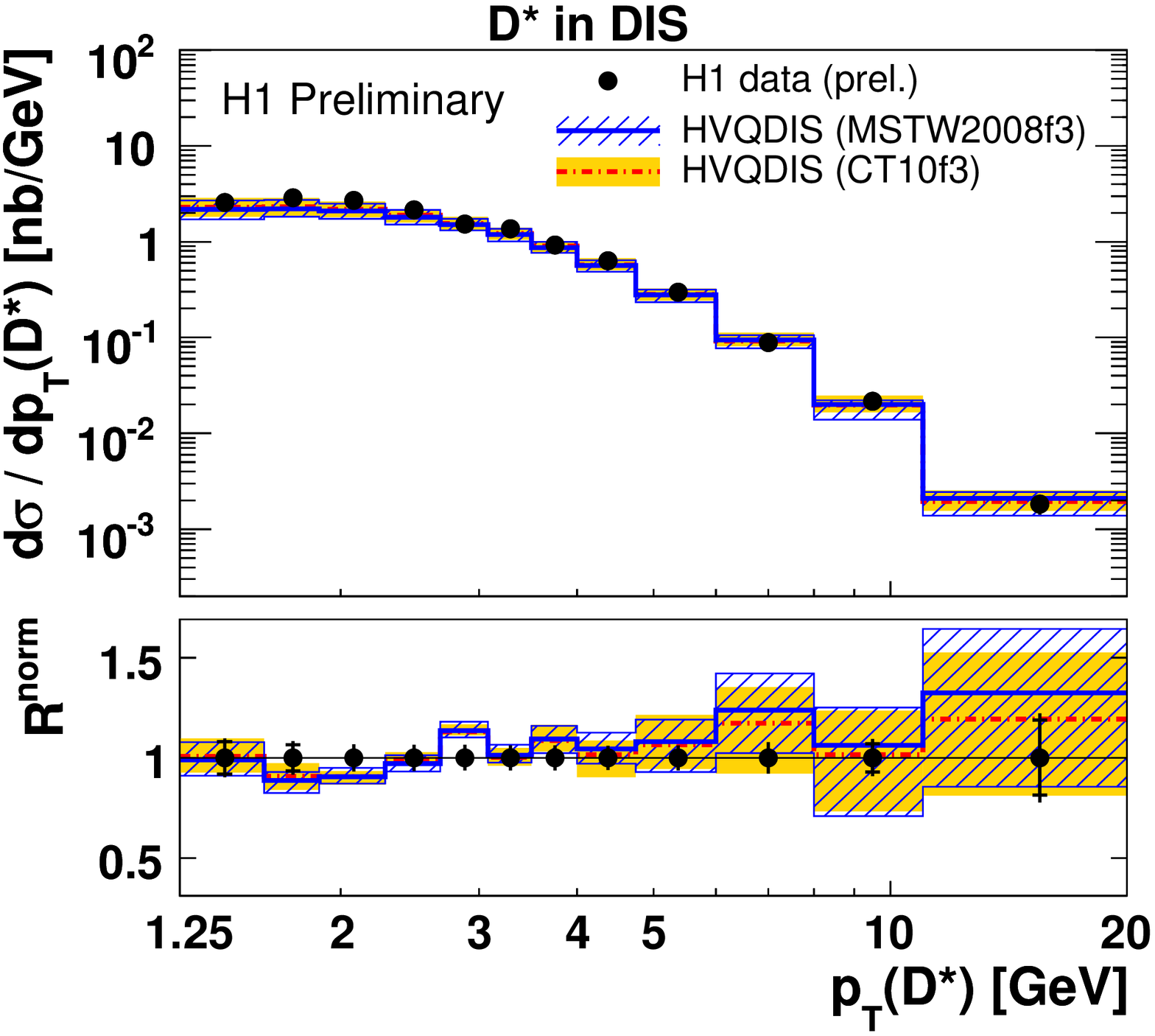}}
     \end{minipage}
     \hspace{.05\linewidth}
     \begin{minipage}[c]{0.475\columnwidth}
    \centerline{\includegraphics[width=0.95\columnwidth]{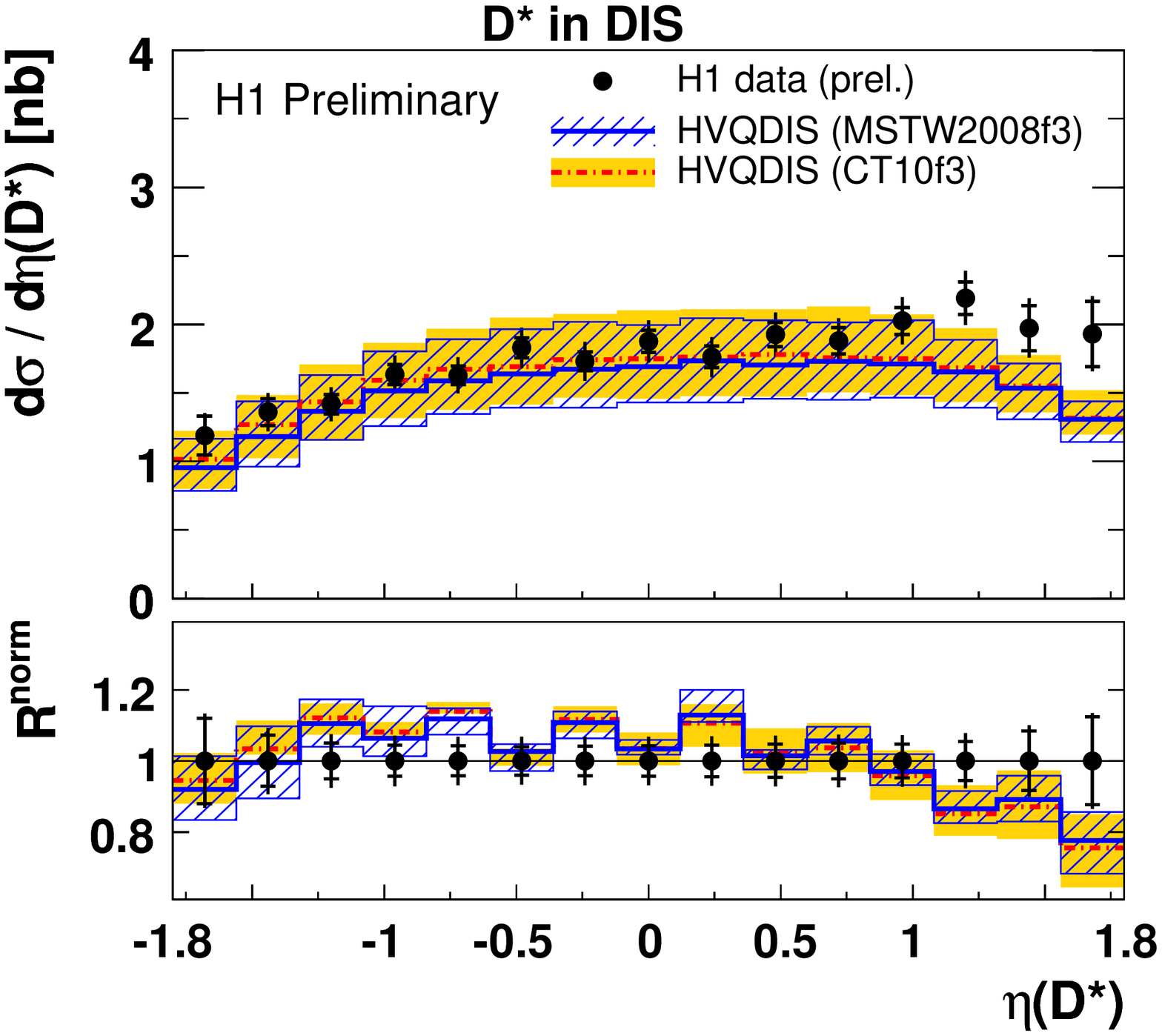}}
     \end{minipage}
  \caption{\label{fig:xsec_pt_eta} $D^*$ cross section as a function of $p_{_{T}}(D^*)$ (left) and $\eta(D^*)$ (right) compared to the massive NLO QCD calculation (HVQDIS) using two different proton PDFs: MRST2008f3 or CT10f3.}    
 \end{figure}
Figure \ref{fig:xsec_pt_eta} shows the $D^*$ cross section as a function of $p_{_{T}}(D^*)$ (left) and $\eta(D^*)$ (right) compared to the massive NLO QCD predictions by HVQDIS. With either proton PDFs the massive NLO calculation describes the data nicely, except for the high $\eta(D^*)$ region (forward direction) where it slightly undershoots the
data. The two-dimensional cross section reveals this to be located at low $p_{_{T}}$ of the $D^*$ (see Figure \ref{fig:xsec_2d}). Otherwise
the data are reasonable described by the NLO calculation.
In order to allow comparisons with the massless NLO calculation (ZM-VFNS) \cite{zmvfns} an additional transverse momentum cut 
in the photon-proton rest frame of $p_{_{T}}^* (D^*) > 2~\mathrm{GeV}$ is applied. The ZM-VFNS calculation uses the CTEQ6.6 proton 
PDF \cite{cteq66} together with the fragmentation function KKKS08 \cite{kkks08}. The massless calculation fails completely to describe the data as a 
function of $x$ (see Fig. \ref{fig:xsec_x_zm}(left)), whereas the data are reasonably well described by the massive NLO QCD calculation 
provided by HVQDIS. Figure \ref{fig:xsec_x_zm}(right) shows the $D^*$ cross section as a function of $y$ which also shows the bad 
description of the data by the massless NLO calculation at low values of $y$.
Differential $D^*$ cross sections have also been measured for a restricted $D^*$ phase space region: $|\eta(D^*)| < 1.5$ and $p_{_{T}}(D^*)>1.5~\mathrm{GeV}$ \cite{medQ2Prel}. Figure \ref{fig:allq2} shows these restricted cross section data as a function of $Q^2$ together with the data from the measurement of $D^*$ cross sections at high $Q^2$: $100 < Q^2 < 1000~\mathrm{GeV^2}$ \cite{highQ2}.
Data are compared to the massive NLO QCD calculation which describes the data well with either proton PDFs, also the $Q^2$ slope is nicely described.

\section{Conclusion}
New measurements using the full H1 HERAII data sample have been analyzed for $D^*$ production in $ep$ scattering. The measurement in the DIS regime at medium $Q^2$ has been carried out in the largest phase space at HERA for inclusive $D^*$ cross section measurements. The data are reasonable described by the massive NLO QCD calculation. However, the NLO calculation undershoots the data slightly at high $\eta(D^*)$ and low $p_{_{T}}(D^*)$ for either PDFs. The slope in $x$ is not very well reproduced for the massive NLO QCD calculation, whereas the massless one fails completely to describe the $x$ and $y$ slopes. A restricted $D^*$ phase space region allows to compare the data from $5 < Q^2 < 1000~\mathrm{GeV^2}$ - this includes the high $Q^2$ $D^*$ data - to the massive NLO QCD predictions which describes the data nicely in slope and normalization.

\smallskip




\begin{thebibliography}{99}

\bibitem{cteq66} P.~M. Nadolsky {\it et al.}, \emph{Phys.~Rev.} {\bf D78} 2008 [{\tt hep-ph/0802.0007}]
\bibitem{medQ2Prel} H1 Collaboration, Measurement of $D^{*\pm}$ Meson Production and Determination of $F_2^{c\bar{c}}$ at low $Q^2$ in Deep-Inelastic Scattering at HERA, 2011 [{\tt hep-ph/1106.1028}].
\bibitem{hvqdis} B.~W. Harris and J.~Smith, Nucl.~Phys.~B {\bf 452} (109) 1995.\\
  Laenen, E. and Riemersma, S. and Smith, J. and van Neerven, W. L., Nucl.~Phys.~B {\bf 392} (162 \& 229) 1993.
\bibitem{mrst08f3} A.D. Martin, W.J. Stirling, R.S. Thorne and G.Watt Eur.~Phys.~J. {\bf C70} (51) 2010 [{\tt arXiv:1007.2624}]
\bibitem{ct10f3} H.~L. Lai {\it et al.}, Phys.~Rev. {\bf D82} 2010 074024 [{\tt arXiv:1007.2241}].
\bibitem{kkks08} T.~Kneesch, B.~A. Kniehl, G.~Kramer and I.~Schienbein, Nucl. Phys. {\bf B799} 2008 [{\tt hep-ph/0712.0481}].
\bibitem{zmvfns} G.~Heinrich, B.~A. Kniehl, Phys.~Rev. {\bf D70} 2004 [{\tt hep-ph/0409303}]; \\
C.~Sandoval, Proc. of XVII International Workshop on Deep-Inelastic Scattering (DIS 2009), Madrid, 2009, [{\tt hep-ph/0908.0824}];\\
C.~Sandoval, Inclusive Production of Hadrons in Neutral and Charged Current Deep Inelastic Scattering, Ph.D. Thesis, Univ. Hamburg (2009), DESY-THESIS-2009-044.
\bibitem{highQ2} F.D. Aaron {\it et al.} [H1 Collaboration], Phys. Lett. {\bf B686} (2010) 91 [arXiv:0911.3989].

\end{thebibliography}
\end{document}